\documentclass[12pt,preprint]{aastex}

\newcommand{\etal}{et~al.\ } 
\newcommand{\PVdblt}{{\rm P}\kern 0.1em{\sc v}~$\lambda\lambda 1117, 1128$} 
\newcommand{\CaIIdblt}{{\rm Ca}\kern 0.1em{\sc ii}~$\lambda\lambda 3934, 3969$}
\newcommand{\AlIIIdblt}{{\rm Al}\kern 0.1em{\sc iii}~$\lambda\lambda 1854, 1862$} 
\newcommand{\CIVdblt}{{\rm C}\kern 0.1em{\sc iv}~$\lambda\lambda 1548, 1550$} 
\newcommand{\MgIIdblt}{{\rm Mg}\kern 0.1em{\sc ii}~$\lambda\lambda 2796, 2803$} 
\newcommand{\NVdblt}{{\rm N}\kern 0.1em{\sc v}~$\lambda\lambda 1238, 1242$}
\newcommand{\SVIdblt}{{\rm S}\kern 0.1em{\sc vi}~$\lambda\lambda 933, 944$} 
\newcommand{\OVIdblt}{{\rm O}\kern 0.1em{\sc vi}~$\lambda\lambda 1031, 1037$} 
\newcommand{\SiIIdblt}{{\rm Si}\kern 0.1em{\sc ii}~$\lambda\lambda 1190, 1193$} 
\newcommand{\SiIVdblt}{{\rm Si}\kern 0.1em{\sc iv}~$\lambda\lambda 1393, 1402$} 
\newcommand{\PV}{\hbox{{\rm P}\kern 0.1em{\sc v}}} 
\newcommand{\AlI}{\hbox{{\rm Al}\kern 0.1em{\sc i}}} 
\newcommand{\AlII}{\hbox{{\rm Al}\kern 0.1em{\sc ii}}}
\newcommand{\AlIII}{{\hbox{\rm Al}\kern 0.1em{\sc iii}}}
\newcommand{\CaII}{\hbox{{\rm Ca}\kern 0.1em{\sc ii}}}
\newcommand{\CII}{\hbox{{\rm C}\kern 0.1em{\sc ii}}}
\newcommand{\CIIe}{\hbox{{\rm C$^{\ast}$}\kern 0.1em{\sc ii}}}
\newcommand{\CIII}{\hbox{{\rm C}\kern 0.1em{\sc iii}}}
\newcommand{\CIV}{\hbox{{\rm C}\kern 0.1em{\sc iv}}}
\newcommand{\CV}{\hbox{{\rm C}\kern 0.1em{\sc v}}}
\newcommand{\HI}{\hbox{{\rm H}\kern 0.1em{\sc i}}}
\newcommand{\HII}{\hbox{{\rm H}\kern 0.1em{\sc ii}}}
\newcommand{\Lya}{\hbox{{\rm Ly}\kern 0.1em$\alpha$}}
\newcommand{\Lyb}{\hbox{{\rm Ly}\kern 0.1em$\beta$}}
\newcommand{\Lyg}{\hbox{{\rm Ly}\kern 0.1em$\gamma$}}
\newcommand{\Lyd}{\hbox{{\rm Ly}\kern 0.1em$\delta$}}
\newcommand{\Lye}{\hbox{{\rm Ly}\kern 0.1em$\epsilon$}}
\newcommand{\Lyphi}{\hbox{{\rm Ly}\kern 0.1em$\phi$}}
\newcommand{\Lyfive}{\hbox{{\rm Ly}\kern 0.1em$5$}}
\newcommand{\Lysix}{\hbox{{\rm Ly}\kern 0.1em$6$}}
\newcommand{\Lyseven}{\hbox{{\rm Ly}\kern 0.1em$7$}}
\newcommand{\Lyeight}{\hbox{{\rm Ly}\kern 0.1em$8$}}
\newcommand{\Lynine}{\hbox{{\rm Ly}\kern 0.1em$9$}}
\newcommand{\Lyten}{\hbox{{\rm Ly}\kern 0.1em$10$}}
\newcommand{\Lyeleven}{\hbox{{\rm Ly}\kern 0.1em$11$}}
\newcommand{\HeI}{\hbox{{\rm He}\kern 0.1em{\sc i}}}
\newcommand{\HeII}{\hbox{{\rm He}\kern 0.1em{\sc ii}}}
\newcommand{\FeI}{\hbox{{\rm Fe}\kern 0.1em{\sc i}}}
\newcommand{\FeII}{\hbox{{\rm Fe}\kern 0.1em{\sc ii}}}
\newcommand{\FeIII}{\hbox{{\rm Fe}\kern 0.1em{\sc iii}}}
\newcommand{\MnII}{\hbox{{\rm Mn}\kern 0.1em{\sc ii}}}
\newcommand{\MgI}{\hbox{{\rm Mg}\kern 0.1em{\sc i}}}
\newcommand{\MgII}{\hbox{{\rm Mg}\kern 0.1em{\sc ii}}}
\newcommand{\MgIII}{\hbox{{\rm Mg}\kern 0.1em{\sc iii}}}
\newcommand{\NI}{\hbox{{\rm N}\kern 0.1em{\sc i}}}
\newcommand{\NII}{\hbox{{\rm N}\kern 0.1em{\sc ii}}}
\newcommand{\NIII}{\hbox{{\rm N}\kern 0.1em{\sc iii}}}
\newcommand{\NV}{\hbox{{\rm N}\kern 0.1em{\sc v}}}
\newcommand{\OVI}{\hbox{{\rm O}\kern 0.1em{\sc vi}}}
\newcommand{\OI}{\hbox{{\rm O}\kern 0.1em{\sc i}}}
\newcommand{\OII}{\hbox{[{\rm O}\kern 0.1em{\sc ii}]}}
\newcommand{\OIV}{\hbox{{\rm O}\kern 0.1em{\sc iv}]}}
\newcommand{\SI}{{\rm S}\kern 0.1em{\sc i}} 
\newcommand{\SIV}{{\rm S}\kern 0.1em{\sc iv}} 
\newcommand{\SVI}{{\rm S}\kern 0.1em{\sc vi}}
\newcommand{\SiI}{\hbox{{\rm Si}\kern 0.1em{\sc i}}}
\newcommand{\SiII}{\hbox{{\rm Si}\kern 0.1em{\sc ii}}}
\newcommand{\SiIII}{\hbox{{\rm Si}\kern 0.1em{\sc iii}}}
\newcommand{\SiIV}{\hbox{{\rm Si}\kern 0.1em{\sc iv}}}
\newcommand{\SII}{\hbox{{\rm S}\kern 0.1em{\sc ii}}}
\newcommand{\SIII}{\hbox{{\rm S}\kern 0.1em{\sc iii}}}
\newcommand{\NaI}{\hbox{{\rm Na}\kern 0.1em{\sc i}}}
\newcommand{\TiII}{\hbox{{\rm Ti}\kern 0.1em{\sc ii}}}
\newcommand{\kms}{\hbox{km~s$^{-1}$}}
\newcommand{\cmsq}{\hbox{cm$^{-2}$}}
\newcommand{\cc}{\hbox{cm$^{-3}$}}
 
\tighten
\begin{document}
 
 


\title{A Quadruple Phase Strong {\MgII} Absorber at $z\sim0.9902$ Toward PG~$1634+706$\altaffilmark{1,2}}

\author{Jie~Ding, Jane~C.~Charlton\altaffilmark{3},
Nicholas~A.~Bond, Stephanie~G.~Zonak, and
Christopher~W.~Churchill\altaffilmark{4}}
\affil{Department of Astronomy and Astrophysics \\ The Pennsylvania
State University \\ University Park, PA 16802 \\ {\it ding, charlton,
bond, szonak, cwc@astro.psu.edu}}

\affil{The Pennsylvania State University, University Park, PA 16802}

\altaffiltext{1}{Based in part on observations obtained at the
W.~M. Keck Observatory, which is operated as a scientific partnership
among Caltech, the University of California, and NASA. The Observatory
was made possible by the generous financial support of the W. M. Keck
Foundation.}
\altaffiltext{2}{Based in part on observations obtained with the
NASA/ESA {\it Hubble Space Telescope}, which is operated by the STScI
for the Association of Universities for Research in Astronomy, Inc.,
under NASA contract NAS5--26555.}
\altaffiltext{3}{Center for Gravitational Physics and Geometry}
\altaffiltext{4}{Visiting Astronomer at the W.~M. Keck Observatory}

\begin{abstract}
The $z=0.9902$ system along the quasar PG $1634+706$ line of sight
is a strong {\MgII} absorber ($W_r (2796) > 0.3$~{\AA}) with only weak
{\CIV} absorption (it is ``{\CIV}--deficient''). To study this system,
we used high--resolution spectra from both the {\it Hubble Space
Telescope} (HST)/Space Telescope Imaging Spectrograph (STIS) and the
Keck I telescope/High Resolution Echelle Spectrometer (HIRES). The STIS
spectrum has a resolution of $R=30,000$ and covers key transitions,
such as {\SiII}, {\CII}, {\SiIII}, {\CIII}, {\SiIV}, and {\CIV}. The
HIRES spectrum, with a resolution of $R=45,000$, covers the {\MgI},
{\MgII} and {\FeII} transitions. Assuming a Haardt and Madau
extragalactic background spectrum, we modeled the system with a
combination of photoionization and collisional ionization. Based on a
comparison of synthetic spectra to the data profiles, we infer the
existence of the following four phases of gas:

i) Seven {\MgII} clouds have sizes of $1$--$1000$~pc and densities of
$0.002$--$0.1$~{\cc}, with a gradual decrease in density from blue to
red. The {\MgII} phase gives rise to most of the {\CIV} absorption and
resembles the warm, ionized inter-cloud medium of the Milky Way;

ii) Instead of arising in the same phase as {\MgII}, {\MgI} is
produced in separate, narrow components with $b \sim 0.75$
{\kms}. These small {\MgI} pockets ($\sim 100$~AU) could represent a
denser phase ($\sim 200$~{\cc}) of the interstellar medium (ISM),
analogous to the small--scale structure observed in the Milky Way ISM;

iii) A ``broad phase'' with a Doppler parameter, $b \sim 60$ {\kms},
is required to consistently fit {\Lya}, {\Lyb}, and the higher--order
Lyman--series lines. A low metallicity ($\log Z \la -2$) for this phase
could explain why the system is ``{\CIV}--deficient'', and also why {\NV}
and {\OVI} are not detected. This phase may be a galactic halo or it
could represent a diffuse medium in an early--type galaxy;

iv) The strong absorption in {\SiIV} relative to {\CIV} could be
produced in an extra, collisionally ionized phase with a temperature
of $T \sim 60,000$~K. The collisional phase could exist in cooling
layers that are shock--heated by supernovae--related processes.
 
\end{abstract}


\section{Introduction}
\label{sec:intro}

Quasar absorption lines provide a unique and intriguing
way to determine the kinematics, chemical composition, and ionization
state of gas in intervening galaxies.  For most galaxies, a
multi--phase interstellar medium (ISM) (i.e. a medium with different
densities and temperatures) is likely to exist
\citep{3phase,multiphase}.  Thus, it follows that any random line of
sight passing through a galaxy is likely to encounter several phases
of gas.

Strong {\MgII} absorbers ($W_r(2796) > 0.3$~{\AA}) are almost always
found within an impact parameter of $35h^{-1}$~kpc of a luminous galaxy
($>0.05L^*$), where $L^*$ is the Schechter luminosity
\citep{bb91,bergeron92,lebrun93,sdp94,s95,3c336}.  In addition, the
strong {\CIV} absorption that is characteristic of most such absorbers
is thought \citep{archiveletter,archive2} to arise in a corona similar
to that surrounding the Milky Way disk \citep{MW}.

We study the $z=0.9902$ system along the line of sight to the quasar
PG~$1634+706$ ($z_{em}=1.335$). This system is a strong {\MgII}
absorber with at least five blended components and with a deficiency
of {\CIV} relative to other systems with similar $W_r({\MgII})$
\citep{archive2}. In a previous study, \citet{stis} derived
constraints on the physical properties of the system, based upon the
combination of low--resolution data from the Faint Object Spectrograph
(FOS) on--board the {\it Hubble Space Telescope} (HST) and
high--resolution data from the High Resolution Echelle Spectrometer (HIRES) on
the Keck I telescope. At that time, only low--resolution data were
available for many key transitions, including low--ionization tracers
({\SiII} and {\CII}), high--ionization tracers ({\SiIV} and {\CIV}),
and the Lyman--series lines. However, the HIRES profiles of {\MgI},
{\MgII}, and {\FeII} were used to form a template that was applied to
model the low--resolution data.

Based on the FOS and HIRES data, \citet{stis} came to the following
conclusions: (1) The metallicity (expressed in the solar units) of the
{\MgII} phase would be low ($-1.5 \la \log Z \la -1$) if it was to
produce a full Lyman limit break; (2) A broad component ($40 \la
b({\rm H}) \la 80$~{\kms}) was needed to fit the {\Lya} profile; (3)
An additional broad component also appeared to be needed to produce
{\CIV}; (4) {\SiIV} could not fully arise in the inferred low-- or
high--ionization phases, which suggested the existence of a blend or
an additional component.

However, some important issues still remained unresolved: (1) whether
the {\MgI} could be produced in the same clouds as {\MgII}; (2)
whether the ionization conditions vary across the profiles; (3)
whether {\CIV} is produced by a single, broad phase or by several
separate, narrower clouds; and (4) whether {\SiIV} is affected by a
blend and, if not, how is it produced?

The first issue is important because {\MgI} is under--produced by the
{\MgII} phase in many strong {\MgII} absorption systems
\citep{thesis,strong2}. In addition, {\MgI} absorption is often very
strong in the environment where damped {\Lya} absorbers (DLAs) arise
and therefore could be used to trace the origin of the strongest {\HI}
absorption. A resolution to the {\MgI}/{\MgII} issue in the case of
the $z=0.9902$ {\MgII} system could serve as a basis for
solving these more general problems.

The third issue is of particular interest because this system is
{\CIV}--deficient \citep{archive2}. Most strong {\MgII} absorbers have
comparable strong {\CIV} absorption, which is likely to arise in a
corona similar to that surrounding the Milky Way. Although {\CIV} is
detected in the low--resolution FOS spectrum, it is relatively
weak. Therefore, the additional component suggested by \citet{stis}
may not be analogous to a corona. The {\CIV} profiles could be
resolved to determine whether the ``corona phase'' is absent, or just
weak in {\CIV}. Either low metallicity or high--ionization conditions
could lead to weak {\CIV} absorption from a corona.

With the release of grating E230M spectra from the Space Telescope
Imaging Spectrograph (STIS) on--board the HST, we now have
high--resolution coverage of the Lyman series, {\SiII}, {\CII},
{\SiIII}, {\CIII}, {\SiIV}, and {\CIV} transitions. In this paper, we
present the results from photo--/collisional ionization modeling of
the $z\simeq0.9902$ {\MgII} system. Combining the newer data with
those on the {\MgI}, {\MgII}, and {\FeII} from Keck/HIRES, we
determine the minimum number of phases required to produce the
observed absorption in the many detected chemical transitions. For
each phase, we place constraints on physical properties such as
densities, temperatures, and sizes of the various environments that
give rise to the absorption features displayed in the observed
spectral profiles.

In \S~\ref{sec:data}, we briefly describe the data that we use to
constrain our models. A summary of our modeling techniques is
presented in \S~\ref{sec:methods} and our major results are outlined
in \S~\ref{sec:results}. Finally, in \S~\ref{sec:discussion}, we
summarize and give a physical interpretation of the system.


\section{Data and Analysis}
\label{sec:data}
\subsection{Keck/HIRES}

The optical part of the spectrum (with {\MgI}~2853, {\MgII}~2796,
{\MgII}~2803 and {\FeII}~2600 covered between 3723~{\AA} and 6186
~{\AA}) was obtained from the HIRES on the Keck I telescope on July 4
and 5, 1995 \citep{Vogt}. This spectrum has a resolution of $R =
45,000$ (FWHM $\sim 6.6$~{\kms}) and a signal--to--noise ratio of
$\sim 50$ per three--pixel resolution element \citep{stis}.
 
The HIRES spectrum was reduced with the IRAF APEXTRACT package for
echelle data and was extracted using the optimal extraction routine of
\citet{horne} and \citet{marsh}. The wavelengths were calibrated to
vacuum using the IRAF task ECIDENTIFY, and shifted to heliocentric
velocities. Continuum fitting was based on the formalism
of \citet{semsav92}. 

\begin{figure*}
\figurenum{1}
\vglue -0.55in
\plotone{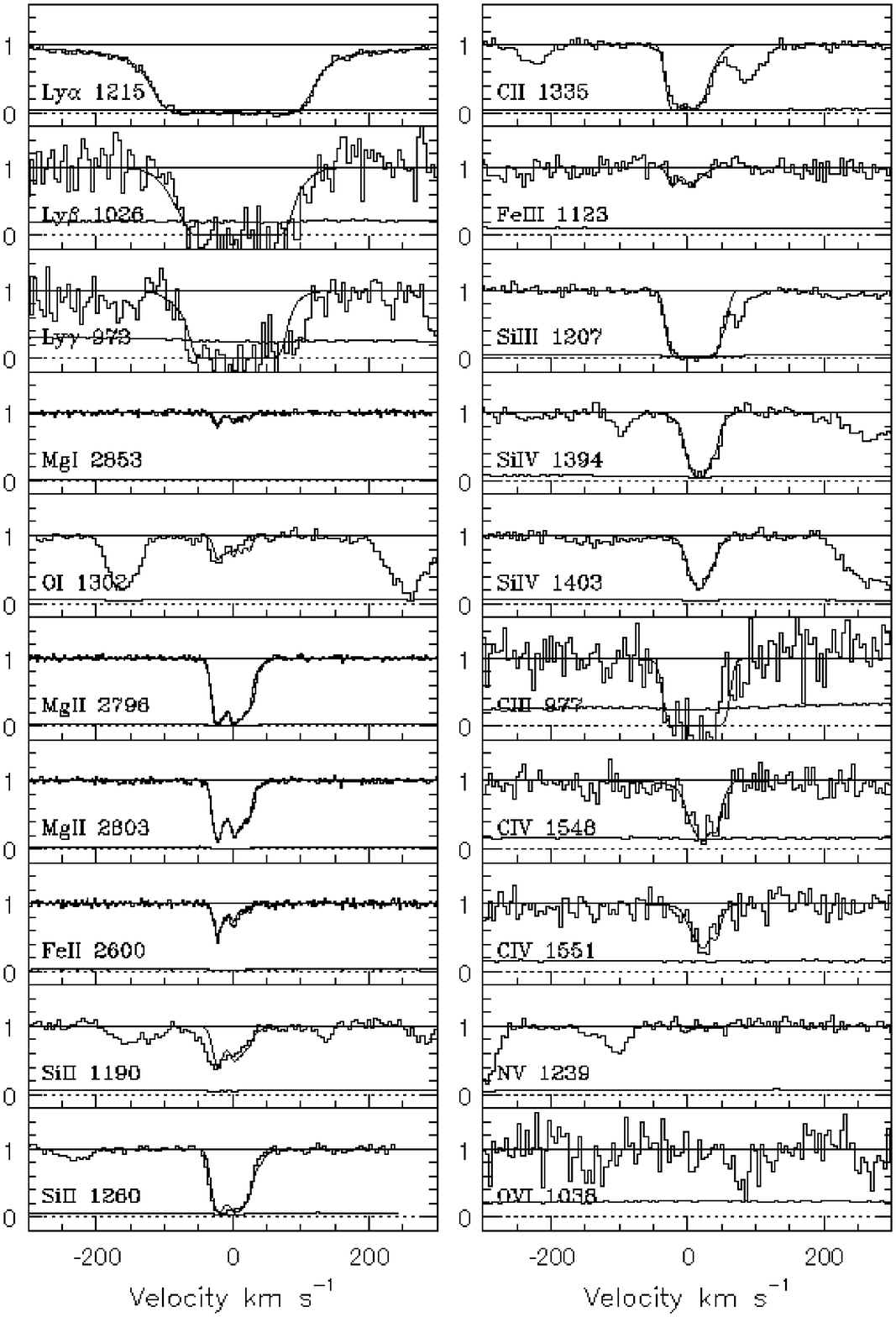} 
\vglue -1.5in
\protect\caption 
{The normalized STIS and HIRES profiles of various key transitions are
displayed in velocity space. The velocity zero--point is at
$z=0.9902$. The solid curve just above zero represents the $1
{\sigma}$ error bar. The curve superimposed on the data is a synthetic
spectrum for one of the best models we have considered (summarized in
Table 1). The {\MgI}, {\MgII}, and {\FeII} profiles were observed with
Keck/HIRES, and all the others with HST/STIS.}
\label{fig:data}
\end{figure*}

Voigt profiles were used to simultaneously fit {\MgI}, {\MgII}, and
{\FeII}, with free parameters redshift, column density and Doppler
parameter, $b$, for each component. The program MINFIT \citep{thesis},
using a $\chi^{2}$ formalism, found the minimum number of components
required to fit the system. The {\MgI}~2853, {\MgIIdblt} and
{\FeII}~2600 transitions, covered by HIRES, are shown in velocity
space in Figure~\ref{fig:data}. Other {\FeII} transitions were also
covered, but {\FeII}~2600 has the highest signal--to--noise.

The fitting procedure was summarized in \citet{cv01}, and described
in detail in \citet{thesis}.  Components are dropped from the fit
unless retaining them produces a $\chi^2$ value that is significantly
better ($97$~\% by the F--test).  Errors in redshifts, column
densities, and Doppler parameters were determined, using the diagonal
elements of the covariance matrix, determined from the curvature
matrix.  Components were rejected if their total fractional errors
exceeded $1.5$.

\subsection{HST/FOS}

Before the release of the STIS data, low--resolution UV spectra were
obtained using the G190H and G270H gratings of HST/FOS.  These spectra
were presented in \citet{impey96} and \citet{kp7}. Since the
higher--resolution STIS spectrum covers most of the key transitions
for the $z=0.9902$ system, we only used the FOS spectra to study the
Lyman limit break at $1830$ {\AA}, which implies the neutral hydrogen
column density $\log N({\HI}) \ga 18$. The data reduction, wavelength
calibration, fluxing, and continuum fitting of the FOS spectra were
performed as part of the QSO Absorption Line Key Project \citep{kp7}.

\subsection{HST/STIS}

Two high--resolution ($R=30,000$) datasets, with different wavelength
coverage, were obtained with HST/STIS and retrieved from the HST data
archive. The E230M spectrum, obtained by S. Burles (May and June 2000),
has a wavelength coverage of $1865$~{\AA} to $2673$~{\AA}, and a total
exposure time of $29,000$ seconds. The E230M spectrum, provided by
B. T. Jannuzi (June 2000), has a wavelength coverage of $2303$~{\AA} to
$3111$~{\AA}, and a total exposure time of $26,435$ seconds. The Lyman
series, {\SiII}, {\CII}, {\SiIII}, {\CIII}, {\SiIV}, and {\CIV} are
all covered by the E230M STIS spectra. A G230M spectrum,
covering 1830~{\AA} to 1870~{\AA}, was also obtained by S. Burles (May
2000), but it did not cover any key transitions for our analysis of
the system.

The STIS spectra were reduced using the standard pipeline
\citep{brown}. The extracted spectra were averaged between different
exposures and the continuum fitting was performed with standard IRAF
tasks \citep{cv01}.

The spectra of various transitions are shown in
Figure~\ref{fig:data}. {\Lyg} is apparently blended with another
feature, based on its asymmetry relative to the other Lyman--series
transitions. {\Lyd} is not shown due to its noisy spectrum and its
contamination by a blend. {\SiII}~1190 is blended with {\NV}~1243 from
the $z=0.9056$ system. {\CII}~1335 is contaminated by {\SiIV}~1394
from the $z=0.9056$ system, but based on {\SiIV}~1403, the
contribution from {\SiIV}~1394 would be negligible. A relatively
strong, unidentified feature is apparent to the red of the {\CII}~1335
profile.  {\SiIII}~1207 is blended with {\SiII}~1260 from the
$z=0.9056$ system. Only {\NV}~1239 is shown because no absorption is
detected in either of the {\NVdblt} profiles and {\NV}~1239 is the
stronger of the two. {\OVI}~1032 is not shown. It is clearly
inconsistent with the undetected {\OVI}~1038.


\section{Methods for Modeling}
\label{sec:methods}

The goal of the present study was to determine the physical conditions
under which the spectra shown in Figure~\ref{fig:data} were produced. To
begin, we applied a Voigt profile fit to the {\MgII} transitions to
determine the locations, column densities, and Doppler parameters of
the components required to reproduce the observed absorption. We
assumed that each one of these components was produced by an
individual ``cloud'' and we modeled each one of these clouds with the
photoionization code Cloudy, version 90.4 (Ferland 1996).

The clouds were assumed to be constant--density, plane--parallel slabs
in photoionization equilibrium. They were ionized by a $z=1$ Haardt
and Madau (1996) extragalactic background spectrum normalizing with an
ionizing photon density of $\log n_{\gamma}=-5.2$.  Alternative input
spectra were also briefly explored, as discussed in
\S~\ref{sec:specshape}. To simulate these clouds, the required input
parameters were: i) the column density of {\MgII}, $N ({\MgII})$, ii)
the metallicity, $Z$ (expressed in the units of the solar value), iii)
the ionization parameter, $U$ (defined as the ratio of the number
density of photons capable of ionizing hydrogen, $n_{\gamma}$, to the
number density of hydrogen, $n_H$), and iv) the abundance pattern. The
column density of {\MgII} was obtained from the Voigt profile
fit. Both the metallicity and the ionization parameter were chosen
from a range reasonable for interstellar and intergalactic clouds
(i.e. $-3 \la \log Z \la 0$ and $-5 \la \log U \la 0$), with the
initial assumption of a solar abundance pattern. If a simultaneous fit
to all transitions was not possible with solar abundances,
alternatives were explored.

Once the input parameters were specified, we ran Cloudy to optimize on
the column density of a selected species, such as {\MgII}. During the
run, Cloudy used an initial guess of $N ({\HI})$ and calculated the
corresponding $N ({\MgII})$. The output $N ({\MgII})$ was compared to
the value measured from the Voigt profile fit, and the value of $N
({\HI})$ was adjusted accordingly for the next iteration. This process
was repeated until the difference between the output $N ({\MgII})$ and
the designated one was negligible.

Upon the completion of the final iteration, the following parameters
were retrieved for the resulting cloud: the column density of every
species with spectral coverage, the size of the cloud, and the
temperature. The temperature was used to calculate the thermal
component of the Doppler parameter of each ion, according to
$b_{therm}^2=2kT/m$, where $m$ is the atomic weight of the ion. The
turbulent component of $b$, which is due to the bulk motion of the gas
and is the same for all the ions, was calculated from the measured $b
({\rm Mg})$ using $b_{turb}^2=b^2({\rm Mg}) - b_{therm}^2({\rm Mg})$. 
Combining the turbulent and thermal components, we obtained the
Doppler parameter of all other individual ions.

Cloudy was run for all five {\MgII} clouds, each time calculating $N$
and $b$ for each species. A ``pure'' spectrum was then produced for
each transition from the $N$ and $b$ of all five components. Then, we
generated a synthetic spectrum by convolving the ``pure'' spectrum
with the instrumental spread function. Finally, the synthetic spectrum
was superimposed on the observed one for comparison.

We experimented with various metallicities and ionization parameters
and attempted to obtain a synthetic spectrum that displayed minimal
discrepancy with the observed one. A visual comparison of goodness of
fit was used to reject models, as illustrated in \citet{stis} and
\citet{weak2}.  This was adequate to confidently constrain parameters
to a relatively narrow range within the very large parameter space.
The $\log U$ and $\log Z$ values given below are accurate to about
$0.1$~dex.  We considered using a quantitative $\chi^2$ approach, but
found it to be impractical because a single pixel or group of pixels
for a particular transition often dominated the assessment.

We encountered several issues that were unresolvable using a single
phase of gas: {\MgI} was under--produced for any choice of $U$; the
shape of the {\Lya} profile could not be fit; the velocity centroids
of {\SiIV} and {\CIV} were offset from that of {\MgII}; and the
{\SiIV} and {\CIV} transitions could not be fit simultaneously.
Additional clouds were included in the {\MgII} phase to account for
the offset {\SiIV} and {\CIV} absorption. The resolution of the other
three issues requires the inclusion of independent phases: A colder,
denser phase is needed to produce the observed absorption in {\MgI}.
A broad, low--metallicity component is needed to account for the shape
of the {\Lya} profile. Finally, a collisionally ionized phase is
needed to fit {\SiIV} without over--producing {\CIV}. Each of these
phases will be explained in detail in \S~\ref{sec:results}.


\section{Results}
\label{sec:results}
Here we present constraints on each of the four phases required to fit
the $z=0.9902$ system, beginning with the lowest--ionization
phase. Table 1 provides a summary of the cloud properties for all four
phases in a plausible model that satisfies all constraints.

\subsection{\MgI~Phase}
\label {sec:MgI}
The five {\MgII} clouds from the original Voigt profile fit could not
produce the observed equivalent width of {\MgI}, regardless of the
choice of ionization parameter.  The ratio $N({\MgI})$/$N({\MgII})$ is
nearly constant at $\simeq 0.01$ over a large range of ionization
parameters, $U$, and for the range of metallicity appropriate for this
system.

\begin{figure*}
\figurenum{2}
\plotone{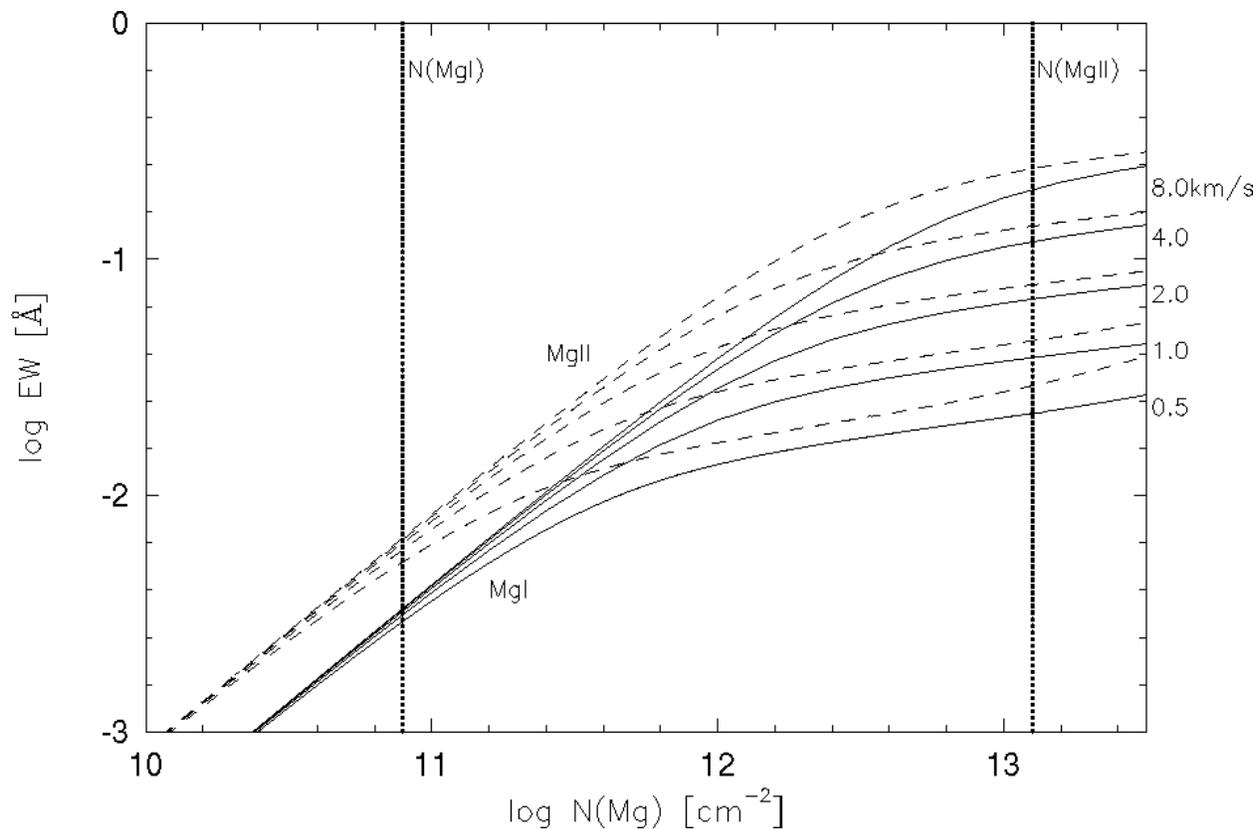}
\protect\caption
{Equivalent width is plotted vs. column density, both on logarithmic
scales. The dashed lines represent curves of growth for {\MgII} 2796,
while the solid ones represent {\MgI} 2853 curves of growth. For both
transitions, a set of Doppler parameters, $b=8.0, 4.0, 2.0, 1.0$ and
$0.5$~{\kms}, are indicated. Dotted lines represent typical
$N({\MgII})$ and $N({\MgI})$ values, i.e. those for the cloud at
$-22$~{\kms}. From the curve it is shown that for our system {\MgI} is
typically on the linear part of the curve of growth, while {\MgII} is
on the flat part.}
\label{fig:Mgcog}
\end{figure*}

In order to obtain the observed equivalent width ratio
$W_r(2853)/W_r(2796)$, a difference in the curve of growth between the
two transitions was used, as shown in Figure~\ref{fig:Mgcog}. In
general, when an absorption profile is unsaturated, its equivalent
width, $W_r$, increases linearly with its column density, $N$ (i.e. on
the linear part of curve of growth). However, when the absorption
becomes saturated, $W_r$ stays almost constant with $N$, at a value
that depends only on Doppler parameter, $b$ (i.e. on the flat part of
curve of growth). In Figure~\ref{fig:Mgcog}, our typical $N({\MgII})$
is on the flat part of the curve of growth, while $N({\MgI})$ is on
the linear part. Thus, we could produce a larger $W_r(2853)/W_r(2796)$
using a smaller $b$, in a ``{\MgI} phase''.

A unique Voigt profile fit to {\MgI} could not be obtained. We assumed
five clouds at the same velocities as the broader {\MgII} cloud
components, with a constant Doppler parameter and a constant
ionization parameter. The Doppler parameter is constrained to be $b
\la 1$~{\kms}, in order that {\MgII} is not over--produced by the
combination of this {\MgI} phase and the {\MgII} cloud phase (see
\S~\ref{sec:MgII}). However, the ionization parameter has to be $\log
U \la -8$ for $b=1$ {\kms}, in order that {\OI} is not
over--produced. Alternatively, $\log U \la -7.5$ if $b=0.75$ {\kms}
({\OI} is on the flat part of the curve of growth).  Column densities
for a sample model with $b({\rm Mg})=0.75$ {\kms} and $\log U=-7.5$ are
listed in Table 1. With $b({\rm Mg})=0.75$ {\kms}, the temperature is
constrained to be lower than $800$~K. Cloudy yields an effective
temperature close to this value for the clouds in our sample model. An
ionization parameter of $\log U=-7.5$ corresponds to a cloud density
of $n(H) \simeq 200$~{\cc}. The cloud sizes are quite small, ranging
from $60$--$160$~AU.

\begin{figure*}
\figurenum{3} 
\plotone{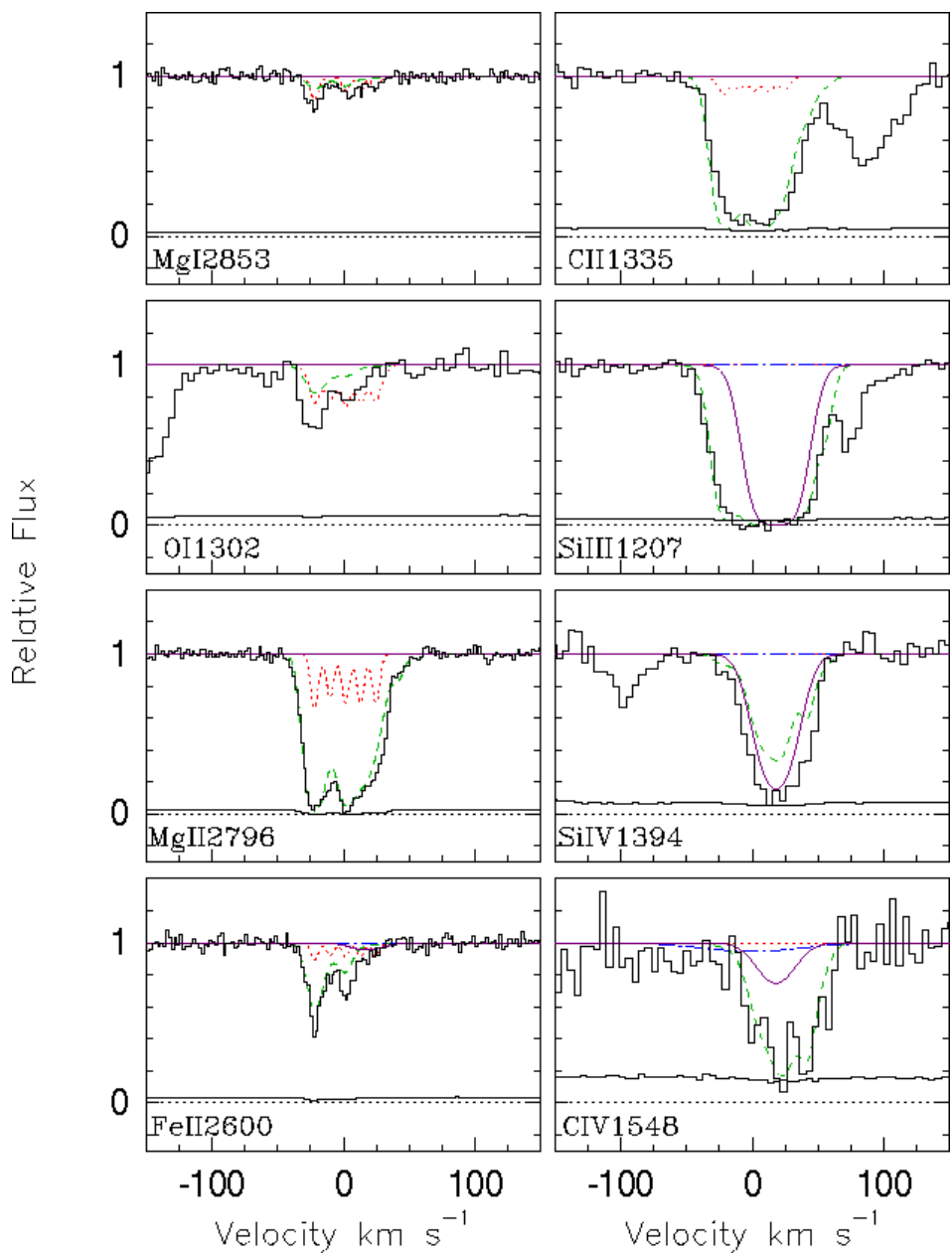} 
\vglue -1.2in
\protect\caption 
{Various key metal transitions are plotted in velocity space. The STIS
and HIRES data are histograms, with model curves superimposed. The
dotted curves represent the {\MgI} clouds. The dashed curves show the
contribution from the {\MgII} phase. The dashed--dotted curves
represent the broad {\HI} phase. The solid curves show the
contribution from the collisional component.}
\label{fig:metals}
\end{figure*}

If the $b$ parameter is small, the {\MgI} phase will have trivial
contribution to the overall equivalent width of any high--ionization
metal transition.  In Figure~\ref{fig:metals}, the dotted lines
represent the contribution from the {\MgI} phase for the model listed
in Table 1. The contribution to {\MgII} is negligible, therefore
{\MgII} has to be entirely produced by the {\MgII} cloud phase (see
\S~\ref{sec:MgII}).

Assuming the five {\MgI} clouds have the same metallicity, the
metallicity of this phase is constrained to be $\log Z \ga -0.8$, by
the {\Lya} profile. At $\log Z=-0.8$, the {\HI} column densities of
the five {\MgI} clouds are in the range of $17.3 \la \log N({\HI}) \la
17.7$, as listed in Table 1, yielding an effective total of $\log
N({\HI}) \simeq 18.1$. Thus, at this metallicity, the {\MgI} phase
will give rise to the full Lyman limit break detected in the FOS
spectrum, which requires $\log N({\HI}) \ga 18$.

\subsection{\MgII~Phase}
\label {sec:MgII}
The column densities and $b$ parameters of the five {\MgII} clouds
were obtained from a Voigt profile fit to the residuals in {\MgII}
unaccounted for by the {\MgI} phase. The five clouds in this phase have
considerably larger $b$ parameters ($6 \la b({\rm Mg}) \la 15$~{\kms}) than
the {\MgI} clouds, as listed in Table 1.

With the detection of strong {\FeII} in clouds ${\MgII}_1$,
${\MgII}_2$, and ${\MgII}_3$, the ionization parameters of these
clouds are constrained by $N ({\FeII})$/$N ({\MgII})$. The remaining
two clouds (${\MgII}_4$ and ${\MgII}_5$) must have higher ionization
parameters, since {\FeII} is very weak. {\SiIV} was used to determine
their ionization states, with the assumption that the {\SiIV}
transitions arise in the same phase as {\MgII}. Thus, the five clouds
have ionization parameters of $\log U \simeq -3.7$, $-4.0$, $-3.5$,
$-3.0$, and $-2.5$, respectively, as listed in Table 1.

The {\SiIV} and {\CIV} profiles are not aligned with the bulk of the
absorption seen in {\MgI}, {\MgII}, and the other low--ionization
transitions.  Thus, additional offset clouds with higher ionization
parameters are needed. A Voigt profile fit to {\SiIV} yielded two
additional clouds (${\MgII}_6$ and ${\MgII}_7$) at $v\sim42~$ and
$v\sim55~${\kms}, respectively. Again, column densities and $b$
parameters of these two clouds are listed in Table 1.  Ionization
parameters are constrained to be $\log U \simeq -2.5$ and $-2.7$, by
$N({\SiIV})$/$N({\CIV})$. This is also consistent with the weak
absorption displayed in the red wings of the {\MgII} and {\SiII}
profiles.

A wide range of ionization states ($-4.0 \la \log U \la -2.5$) is
displayed in this {\MgII} phase, with a tendency for increase in
ionization stage from blue to red in the spectrum. Therefore, most of
the absorption in both low--ionization transitions, such as {\MgII},
{\FeII}, {\SiII}, and {\CII}, and high--ionization transitions, such
as {\SiIII}, {\CIII}, and {\CIV}, is produced in this phase, as shown
in Figure~\ref{fig:metals}. However, the substantial under--production
of {\SiIV} suggests the existence of an extra component, as discussed
in \S~\ref{sec:collisional}.

Assuming that all seven clouds in the {\MgII} phase have the same
metallicity, the metallicity is constrained to be $\log Z \ga -0.65$
by the Lyman series. However, even at the lower limit of $\log Z =
-0.65$, with $14.9 \la \log N({\HI}) \la 17.1$, the {\MgII} phase
contributes insignificantly to the observed full Lyman limit
break. Alternatively, if the {\MgI} and {\MgII} phases are assumed to
have the same metallicities, a metallicity of $\log Z \simeq -0.8$ is
required by the {\Lya} profile (Figure~\ref{fig:hydro}).  As mentioned
in \S~\ref{sec:MgI}, at this metallicity the five {\MgI} clouds will
give rise to the observed full Lyman limit break (see Table 1).

\subsection{Broad \HI~Phase}
\label {sec:HI}
\begin{figure*}
\figurenum{4}
\vglue -0.5in
\plotone{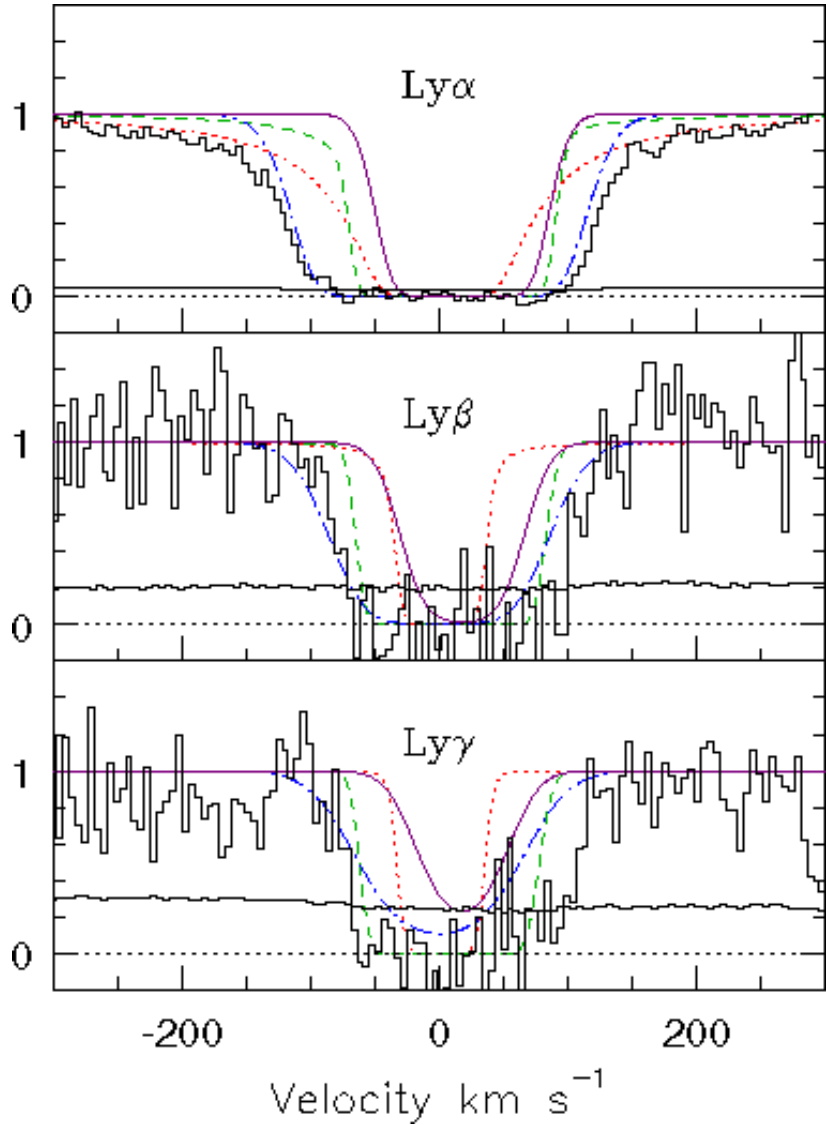}
\vglue -1.6in
\protect\caption
{The {\Lya}, {\Lyb}, and {\Lyg} profiles are displayed in velocity
space. The solid histograms represent the STIS data. Other curves show
the results for our ``best model'' (listed in Table 1). The model
curves represent different phases as indicated in the caption of
Figure~\ref{fig:metals}.}
\label{fig:hydro}
\end{figure*}

The {\Lya} profile is too broad to be fully produced by the {\MgI} and
{\MgII} clouds, as is shown in Figure~\ref{fig:hydro}. However, if
the metallicity was decreased to give rise to more {\Lya}, the
narrower {\Lyg} would be over--produced. The curve of growth of the
Lyman--series lines \citep{q1206} allows for a centered, broad
component that can resolve this dilemma (i.e. it accounts for the broader,
deeper {\Lya} profile, as well as the wing structure in {\Lyb},
without over--producing the narrower {\Lyg} profile.)

A Voigt profile fit to {\Lya} yielded the column density and $b$
parameter of the broad phase. If $b({\rm H})\ga 70$~{\kms} and
$N({\HI})\la 10^{15}~${\cmsq}, we can produce the saturated absorption
from $-100$ to $+120$~{\kms}, but the wings will be over--produced; if
$b({\rm H}) \la 50$~{\kms} and $N({\HI}) \ga 10^{16}~${\cmsq}, we can
produce the range of the saturated absorption, but the wings will be
under--produced. The best fit was found to be with $N({\HI}) \sim
10^{15.5}~${\cmsq} and $b({\rm H}) \sim 58$~{\kms}.

In principle, the broad component could be either photoionized or
collisionally ionized. In both cases, the metallicity is constrained so
as not to over--produce any of the high--ionization transitions
(i.e. {\CIV}, {\NV} and {\OVI}).  The {\NV} and {\OVI} are not
detected, and {\CIV} is fully produced in the {\MgII} phase (as
required by the structure appearing in the {\CIV} profiles.).

For photoionization, an upper limit of $\log Z \la -2$ is placed on
the metallicity, regardless of the choice of ionization parameter.
Otherwise, either low-- or high--ionization transitions would be
over--produced.  If the metallicity is $\log Z \la -2.5$, there will
be no constraint on the ionization parameter from the metal
transitions. None of them is significantly produced at such a low
metallicity. However, a rough upper limit of $\log U \la -1$ can be
estimated by the requirement of a reasonable cloud size (less than
several hundred kpcs).  A good fit for this phase was obtained for a
model with a metallicity of $\log Z \simeq -2.5$ and an ionization
parameter of $\log U \simeq -1.5$, as listed in Table 1.  This model
gives a cloud $\sim 60$~kpc in size. The fit to the Lyman series is
shown by the dashed--dotted curve in Figure~\ref{fig:hydro}.

For collisional ionization with pure thermal broadening (i.e. no bulk
motion), the temperature would be $T \sim 10^{5.3}$~K, using $b^{2} =
2kT$/$m$ for {\HI}. At this temperature, the metallicity has to be
exceptionally low, $\log Z \la -3.5$, in order not to over--produce
the high--ionization transitions. If, instead, bulk motion
contributes, the $b$ parameter will be larger for each metal species
than the one given by pure thermal scaling of $b({\rm H})$. For
example, as the temperature decreases from $T \sim 10^{5.3}$~K, $N
({\CIV})$ increases towards a peak at $T \sim 10^{5.0}$~K. This,
together with a larger $b({\rm C})$, gives a larger {\CIV} equivalent
width when bulk motion is involved. In order not to over--produce
{\CIV}, the metallicity has to be even lower than the pure thermal
case. A metallicity as low as $\log Z \sim -3.5$ is
unlikely. Therefore, collisional ionization is ruled out for this
phase. The broad {\HI} absorption is more likely to arise in a
photoionized cloud, but even then the metallicity is constrained to be
quite low ($\log Z \la -2$).

\subsection{Collisional Ionization Phase}
\label {sec:collisional}

The broad, smooth absorption profiles in the {\SiIV} transitions are
not consistent with the structure apparent in the {\CIV}
profiles. Because {\SiIV} is offset relative to {\Lya}, it also cannot
be produced by the centered, broad {\HI} phase. The need to produce
{\SiIV} without contributing a substantial amount of absorption to
other transitions suggests the existence of a collisionally ionized
phase. A Voigt profile fit to the residuals in {\SiIV} yielded a
component at $v \sim 17~${\kms}, with $N({\SiIV}) \sim
10^{13.5}~${\cmsq} and $b({\rm Si}) \sim 17$~{\kms}.  However, for
collisional ionization, {\SiIV} peaks at $\log T \sim 4.8$
\citep{cooling function}, which corresponds to $b_{therm} \sim 6
$~{\kms}. Therefore, bulk motion must also contribute to the overall
velocity dispersion. The temperature range was found to be $10^{4.7}
\la T \la 10^{4.9}$~K, in order not to over--produce the transitions
of higher-- and lower--ionization stages. The metallicity of this
phase could be set to the same value as the {\MgI} and {\MgII} phases,
$\log Z \simeq -0.8$, but it is not well constrained. A lower limit of
$\log Z \ga -3$ is placed by the the Lyman--series lines.

\subsection{Effects of Assumed Input Spectrum}
\label{sec:specshape}

For all the previous results, we have used the $z=1$ Haardt and Madau
extragalactic background radiation (EBR) spectrum
\citep{haardtmadau96} as the only source of ionizing photons.  Here we
consider the likelihood of additional stellar contributions and
consider the effect of such alternative spectra in a couple of extreme
examples.

At redshift $z\sim1$ the extragalactic background radiation is more
likely to dominate stellar contributions than it is at other
redshifts.  At lower redshifts, the amplitude of the Haardt and Madau
extragalactic background is lower (by a factor of seven at $z=0$
\citep{haardtmadau96}).  At much higher redshifts, starbursts are
common and the amplitude of the EBR from quasars has levelled off.

Unfortunately, in the case of the PG~$1634+459$ line of sight, we have
no information about the colors or morphologies of the host galaxies.
Therefore, we cannot rule out a starburst host for the $z=0.9902$
absorber.  Only an extreme starburst would affect the results, and
only within $\sim 6$~kpc of the center of such a galaxy would the
stellar contribution strongly dominate the EBR ($1$\% of the photon
flux of $10^{54}~{\rm s}^{-1}$ would escape the burst region
\citep{starburst}).  We would expect the absorption from the central
region of a starburst to be stronger than it is for this $z=0.9902$
system \citep{bond}.  Furthermore, we would expect to pass through
some layers of highly ionized gas, and this absorber is weak in {\CIV}
and undetected in {\NV} and {\OVI}.  Therefore, a strong starburst
host galaxy is unlikely, but we still consider here the effect of a
couple of modified spectra.

Many different spectral shapes would be possible once we allow a
stellar contribution and it is unrealistic to consider them all, so we
choose two extreme cases: 1) a $0.01$~Gyr instantaneous burst, which
is characterized by the edges of {\HI}, {\HeI}, and {\HeII}; and 2) a
$0.1$~Gyr instantaneous burst with an extreme {\HI} edge, but with a
relatively flat spectrum above that edge, similar to the Haardt and
Madau shape.  Both stellar spectra, assuming solar metallicity and a
Salpeter IMF, were taken from \citet{bruzual}. We have run models with
the burst contribution set to ten times that of the EBR, at $1$
Rydberg.  We begin with the model in Table~1 and consider how the
addition of the burst would affect our conclusions. In a more
realistic model, however, the spectral shape would be a function of
position, with the EBR likely to be dominant at some locations, but
the solution to that case would be in between the extreme starburst
models and a pure Haardt and Madau model.

Stellar contributions from a non-burst model are more likely to
dominate the EBR at energies less than $1$~Rydberg.  This could affect
the ionization balance between the neutral and singly ionized
transitions.  Because of its extreme Lyman limit break (almost a
factor of $1000$), the $0.1$~Gyr instantaneous burst model can also be
used to represent the rough effects in this case.

For the $0.01$~Gyr model, due to the softer spectrum (because of the
edges of helium), the {\CIV} is under--produced compared to the pure
Haardt and Madau case.  For the same reason, for the clouds optimized
on {\SiIV}, the lower--ionization transitions are found to be
over--produced compared to {\SiIV}.  With this spectral shape, we
cannot produce even the weak, observed {\CIV}.  In principle, this
could relate to why this system is {\CIV}--deficient, but in practice
it is difficult to understand how such a strong stellar flux could affect a
large enough region to reduce the high--ionization absorption at all
velocities.  The metallicity that we would infer for the $0.01$~Gyr
model would be quite similar (within $0.1$~dex) to that for the pure
Haardt and Madau case.

For the $0.1$~Gyr model, there are relatively fewer ionizing photons
just above the {\HI} edge, and so more hydrogen remains neutral.  This
gives rise to a stronger {\Lya} line.  To match the observed {\Lya}
profile, the metallicity would have to be increased, but only by $\sim
0.3$~dex.  Due to the similar shape to the Haardt and Madau spectrum
above the edge, the ionization conditions are not very different for
metal--line transitions.  Slightly more {\CIV} is produced in the
$0.1$~Gyr model, but this would require an ionization parameter
adjustment of only $\sim 0.1$~dex.  It is also important to note that
for neither burst model is our conclusion altered about the need for
an additional {\MgI} phase.  However, the $0.1$~Gyr burst model does
give rise to a larger amount of {\OI} in the {\MgI} clouds, which
would somewhat change constraints on the cloud densities and sizes.


\section{Summary and Discussion}
\label{sec:discussion}

Based upon modeling of high--resolution HST/STIS and Keck/HIRES
absorption profiles of the $z=0.9902$ strong {\MgII} absorber
toward PG~$1634+706$, we derived the physical conditions of the phases
of gas encountered along the line of sight.  Here, we summarize the
physical conditions (densities, temperatures, size of structures) that
characterize each of the phases.  These properties indicate
relationships between the gas phases and various structures in the
Milky Way, as well as in other galaxies in the local universe.

\subsection{{\MgII} Clouds}

Seven blended components with $b$ parameters ranging from
$4$--$15$~{\kms} provide a consistent fit to the {\MgIIdblt} profiles,
which extend over $\sim 75$~{\kms} in velocity. Under the assumption
of a Haardt and Madau ionizing spectrum, these photoionized clouds have
densities of $0.002$--$0.1$~{\cc} (corresponding to $-4.0 \la \log U
\la -2.5$), with a gradual decrease from blue to red, and temperatures
of $\sim 10,000$~K.  {\CIV} was also fully produced in this phase. We
infer a metallicity of $\log Z \simeq -0.8$, assuming that these
clouds have the same metallicity as those of a separate {\MgI} phase.
The cloud sizes/thicknesses range from parsecs to hundreds of parsecs,
as listed in Table 1.

The physical conditions in this phase are similar to those in the
warm, ionized inter-cloud medium of the Milky Way \citep{3phase}. They
are also similar to most other {\MgII} clouds in strong {\MgII}
absorbers \citep{archive2}.  The similarity with the Milky Way disk
ISM and the connection between strong {\MgII} absorbers and the
majority of $L^*$ galaxies suggests that the {\MgII} absorption comes
from the warm ISM of a spiral disk.  However, strong {\MgII}
absorption can also come from early--type galaxies \citep{csv96}. In
fact, \citet{archive2} found that the three {\CIV}--deficient {\MgII}
absorbers in their sample had among the reddest $B-K$ colors and the
highest luminosities.  The similarity of the low--ionization gas
properties between ``classic'' strong {\MgII} absorbers and
{\CIV}--deficient {\MgII} absorbers would then suggest that
early--type galaxies house regions with an ISM similar to that in
spirals.

\subsection{Pockets of {\MgI}}

Perhaps the most interesting and surprising result of our modeling was
our inference of the existence of a cold phase ($T<1000$~K).  It
consists of dense ($n({\rm{H}}) \sim 200$~{\cc}) pockets that give
rise to the bulk of the {\MgI} absorption in the form of very narrow
($b<1$~{\kms}) components.  This phase also gives rise to the observed
full Lyman limit break if $\log Z=-0.8$ for all the clouds.  These
``{\MgI} pockets'' are quite small, only $\sim 100$~AU.

From many different techniques, there is conclusive evidence for the
existence of small--scale structure in the Milky Way ISM down to scales
of $100$~AU and lower.  The techniques include high--resolution
absorption studies toward globular clusters and binary stars
(eg. \citet{andrews01}, \citet{meyer99}, \citet{watson96}, and
\citet{meyer96}) and {\HI} 21-cm absorption studies toward extragalactic
sources and high--velocity pulsars (eg. \citet{faison98} and
\citet{frail94}).  The highest--resolution absorption studies yielded
small $b$ parameters, $1$--$2$~{\kms}, for some of the features
\citep{andrews01}, consistent with our proposed {\MgI} pockets.

The observed small, high--density structures have been difficult to
reconcile with the idea that the ISM is in pressure balance.  Their
pressures are two orders of magnitude larger than the pressure of the
warm, ionized inter-cloud medium.  In the context of this $z=0.9902$
system, the pressure of these {\MgI} pockets would be inconsistent
with pressure confinement by the {\MgII} clouds.

There have been several efforts to reconcile the issue of pressure
balance in the interstellar medium.  \citet{elmegreen} has
investigated a fractal structure model, driven by turbulence, that
leads to significant clumping on small scales.  \citet{heiles} has
considered modifications of the standard cooling paradigms as well as
alternative geometries.  \citet{walker} investigated the idea that the
small clouds are self--gravitating.  Regardless of the mechanism of
formation and stability of these small structures, it is clear that
they are pervasive in the disk of the Milky Way.  Therefore, it is not
surprising that we would see them in absorption through other galaxies
such as this one probed by the PG~$1634+706$ line of sight.  We should
note, however, that since there is no definite theory for their
existence, our assumption that these clouds are in equilibrium may
only be an approximation to a more complex physical situation.

A clue about the spatial arrangement of the cold and warm absorbing
gas in the $z=0.9902$ system comes from the relative sizes we have
derived from the photoionization models.  The sizes of the {\MgI}
pockets are $10^3$--$10^5$ times smaller than those of the {\MgII}
clouds, yet we find several of them along this line of sight.  This
suggests a sheet--like structure such that the covering factor is
large, or perhaps a fractal structure as proposed by
\citet{elmegreen}. In addition, depending on the geometry, such small
{\MgI} clouds could only partially cover the quasar beam, a phenomenon
generally associated with absorbing clouds intrinsic to the quasar
(e.g. \citet{barlow}, \citet{hamann}, \citet{ganguly}, and
\citet{dekool}).  The quasar continuum source beam size would be
$\sim10$--$100$ Schwarszchild radii at the distance of the quasar,
$\sim20$--$200$~AU for a $10^8$~$M_{\odot}$ black hole.

In fact, the large $W({\MgI})/W({\MgII})$ ratio that led to our
proposal of the existence of the {\MgI} cloud phase is common in
strong {\MgII} systems.  In about half the strong {\MgII} clouds, a
simultaneous Voigt profile fit to the {\MgI} and {\MgII} profiles
yielded $N({\MgI})/N({\MgII})$ larger than could be reconciled with a
single--phase photoionization model \citep{thesis,strong2}.  Also,
\citet{rauch} found a large {\MgI} column density for the strong
{\MgII} system at $z=0.5656$ toward Q~$2237+0305$, and found that it
implied either an extremely large density, or a large $N({\HI})$.
They favored the latter, suggesting that this is a strong Lyman limit
or damped {\Lya} system.  Our results here imply that the former
solution, of high--density pockets, is also reasonable.

The existence of these tiny ``{\MgI} pockets'' may also provide a hint
about the nature of damped {\Lya} absorbers, especially those at
relatively low redshifts.  Unlike strong {\MgII} absorbers, which
almost always have a $L>0.05 L^*$ galaxy within $40 h^{-1}$~kpc
\citep{bb91,bergeron92,lebrun93,sdp94,s95,3c336}, DLAs have a variety
of types of galaxy hosts \citep{rao00}, including dwarfs and low-
surface-brightness galaxies (LSBGs)
\citep{3c336,turnshek01,kulkarni01,bowen01,bouche01}.  If the
strongest {\HI} absorption (the DLAs) occurs in a continuous medium in
the central region of a galaxy, it seems difficult to explain why
Lyman limit absorption ($17.2 < \log N({\HI}) < 20.3$) would not occur
in a relatively large annulus surrounding the central regions of
dwarfs and LSBGs.  However, if the DLA is produced in small,
high--density pockets of material, it is plausible that the surrounding
regions could be evacuated of gas in galaxies with small potential
wells.  Thus, dwarfs and LSBGs could give rise to DLA absorption but
not to a significant fraction of ordinary Lyman limit systems.

The idea that DLAs come from small pockets is supported by other
observations.  \citet{lane00} have resolved narrow components
($2$--$3$~{\kms}) in an {\HI} 21-cm absorption profile of the DLA at
$z=0.0912$ toward B~0738+313.  With some thermal scaling, it would be
expected that the associated {\MgI} components would be even narrower,
consistent with what we have found in our $z=0.9902$ absorber.
Furthermore, the DLAs with larger {\MgI} equivalent widths tend to be
the ones with many components in their 21-cm absorption profiles (Rao
2002; private communication), and many of these components are also
quite narrow \citep{lane00}.  Finally, the high--ionization absorption
associated with DLAs is similar to that of classic strong {\MgII}
absorbers, as if it is due to parts of the galaxy that are unrelated
to the DLA region \citep{archive2}.  DLAs could simply be small
concentrations of gas that exist in large numbers in many kinds of
environments.  Our strong {\MgII} absorber at $z=0.9902$ could be a
less extreme (in its $N({\HI})$) and more common version of the same
sort of structure.

\subsection{Alternatives to Separate {\MgI} and {\MgII} Phases}
\label{sec:alternatives}

In our favored model, the {\MgII} phase (with $6 < b({\rm Mg}) <
15$~{\kms}) severely underproduces {\MgI}, while the {\MgI} phase
(with $b \sim 0.75$~{\kms}) produces only negligible {\MgII}.  This
might suggest that there would be an alternative model with
intermediate $b$ values so that {\MgI} and {\MgII} can be produced in
the same clouds.  However, we find that this is not the case.
Although, with a smaller value of $b$ the observed equivalent width
ratio of {\MgI} to {\MgII} can be matched, the fit to {\MgII} is
unsatisfactory (too deep and too narrow).

Alternatively, we could relax our assumption that the {\MgII} profile
is fit with the minimum number of component clouds that produce an
adequate fit to the data.  A larger number of clouds could be used,
with smaller $b$ values (perhaps $2$--$3$~{\kms}).  It would seem that
we could increase the number of clouds (thus decreasing the maximum
$N({\MgII})$ for a cloud) so that {\MgII} is no longer on the flat
part of its curve of growth.  With {\MgII} on the linear part of its
curve of growth, $W({\MgI})/W({\MgII})$ would be increased.  In fact,
this is not an acceptable solution for this system.  As the number of
clouds is increased, there is a competing effect that reduces
$W({\MgI})/W({\MgII})$.  Increased blending of the superimposed clouds
affects the {\MgI} more significantly than the {\MgII}.  This effect
dominates so that we are unable to find a model with many moderate $b$
clouds to fit the data.

We conclude that a two--phase model, with separate phases producing
the {\MgI} and {\MgII} absorption, is the simplest suitable model that
can fit both the {\MgI} and {\MgII} profiles and those of other
intermediate ionization transitions.

\subsection{Broad {\HI} phase}

A broad {\HI} component, with $b\sim60$~{\kms} was proposed to
produce a self--consistent fit to the {\Lya}, {\Lyb}, and {\Lyg}
profiles.  In order that this component does not over--produce
any of the high--ionization metal--line transitions, it must have
a low metallicity, $\log Z \la -2$.  It is consistent with a
photoionization model, but the ionization parameter is poorly
constrained (since this phase does not give rise to metals).
The size of this cloud is therefore also poorly constrained.
If $\log U=-2.5$ the size would be $\sim 60$~kpc.

In most other strong {\MgII} absorbers, the {\CIV} absorption is too
strong, and the components too broad, for it to arise in the same
phase as the {\MgII} \citep{archive2}.  These absorbers also typically
require an additional phase to self--consistently fit the Lyman
series, and it is possible to produce both the high--ionization
absorption and the additional {\HI} absorption with the same phase.
The need for a broad, high--ionization phase is especially evident in
cases like the ``double'' strong {\MgII} absorber at $z=0.9276$ toward
PG~$1206+459$, in which {\CIV}, {\NV}, and {\OVI} absorption are all
very strong \citep{q1206,q1206new}.  Such strong, broad,
high--ionization absorption also characterizes the corona that
surrounds the disk of the Milky Way \citep{MW,savage00}.

The broad {\HI} phase of the $z=0.9902$ absorber resembles the {\HI}
component of the broad phases proposed for those other strong {\MgII}
absorbers.  However, in this case no high--ionization metal--line
absorption is produced by the broad phase, which led us to the
conclusion that it has very low metallicity.  With such a low
metallicity $\log Z \la -2$ as compared to its {\MgII} clouds ($\log Z
\sim -0.8$), it seems unlikely that this system is analogous to the
Milky Way corona.  In fact, the low metallicity is more suggestive to
a halo structure.  The $b$ parameter of $\sim 60$~{\kms} could also be
consistent with such an interpretation.

This leads back to why the system is ``{\CIV}--deficient'' (and {\NV}
and {\OVI}--deficient as well).  It is more likely that this absorber
is produced by a galaxy without a significant corona. The alternative,
a low--metallicity corona, is less likely since a relatively high
metallicity ($Z \simeq -0.8$ for the {\MgI} and {\MgII} phases) would
be indicated for the disk that would produce the corona.  An
early--type galaxy would be consistent with this interpretation, but is
certainly not a unique solution.  It may be only a coincidence that
the broad {\HI} component in this {\CIV}--deficient absorber resembles
those needed to fit the Lyman series in most other strong {\MgII}
absorbers.

\subsection{Collisionally Ionized Phase}

Another unusual feature of the $z=0.9902$ system is the relatively
strong, smooth {\SiIV} profile, fit with $b \sim 17$~{\kms}.  Since the
cloud structure apparent in the {\CIV} profile matches the {\MgII}
clouds, we infer collisional ionization with a temperature close
to the peak for {\SiIV} production, $T \sim 60,000$~K.

It is interesting to note that a similar situation was found in the
case of the $z=1.04$ multi--cloud weak {\MgII} absorber along this
same quasar line of sight \citep{zonak}.  In that case, the {\SiIII}
profile was smooth and featureless, and also much stronger than
expected from photoionization models of the other phases inferred for
that system.  In that case, a somewhat smaller temperature of
$40,000$~K provided an adequate fit.

The temperatures of these additional, collisionally ionized phases
place them on an efficient part of the cooling curve, so they would
have to be quite common to be detected during a relatively brief
interval.  However, it is reasonable to think that cooling layers
heated by supernovae shocks could provide an appropriate environment.
More complex models than pure collisional ionization have been
considered to explain the ratios of high--ionization metal--line
transitions in the Milky Way disk \citep{MW}.  Certainly, more complex
processes than pure collisional ionization could be involved in this
case as well.  If strong, smooth profiles for one particular
transition are commonly found in strong {\MgII} absorbers, they could
be compared in more detail to supernova models and used to track the
evolution of the ensemble of supernovae remnants in the universe.


\clearpage

\clearpage

\newpage
\begin{deluxetable}{lrcclrcccccccccccc}
\label{tab:tab1}
\tablenum{1}

\tabletypesize{\scriptsize}
\rotate
\tablewidth{0pt}
\tablecaption{Cloud Properties}
\tablehead{
\colhead{} &
\colhead{$v$}  &
\colhead{$Z$} &
\colhead{$\log U$} &
\colhead{$n_H$} &
\colhead{size} &
\colhead{$T$} &
\colhead{$N_{\rm tot}({\rm H})$} &
\colhead{$N({\HI})$} &
\colhead{$N({\MgI})$} &
\colhead{$N({\MgII})$} &
\colhead{$N({\FeII})$} &
\colhead{$N({\SiIV})$} &
\colhead{$N({\CIV})$} &
\colhead{$b({\rm H})$} &
\colhead{$b({\rm Mg})$} &
\colhead{$b({\rm Fe})$} \\
\colhead{} &
\colhead{[{\kms}]} &
\colhead{[$Z_{\odot}$]} &
\colhead{} &
\colhead{[{\cc}]} &
\colhead{[pc]} &
\colhead{[K]} &
\colhead{[{\cmsq}]} &
\colhead{[{\cmsq}]} &
\colhead{[{\cmsq}]} &
\colhead{[{\cmsq}]} &
\colhead{[{\cmsq}]} &
\colhead{[{\cmsq}]} &
\colhead{[{\cmsq}]} &
\colhead{[{\kms}]} &
\colhead{[{\kms}]} &
\colhead{[{\kms}]}
}

\startdata
\multicolumn{15}{c}{\sc {\MgI}~Phase}\\
\hline
{\MgI}$_{\rm 1}$ & $-22$ & $0.16$ & $-7.5$ & $200$ & $0.0008$ & $500$ & $17.7$ & $17.7$ & $11.1$ & $12.4$ & $11.9$ & $0$ & $0$ & $3.7$ & $0.75$ & $0.5$ \\
{\MgI}$_{\rm 2}$ & $-10$ & $0.16$ & $-7.5$ & $200$ & $0.0003$ & $700$ & $17.3$ & $17.3$ & $10.4$ & $12.1$ & $11.7$ & $0$ & $0$ & $3.7$ & $0.75$ & $0.5$ \\
{\MgI}$_{\rm 3}$ &   $1$ & $0.16$  & $-7.5$ & $200$ & $0.0005$ & $600$ & $17.5$ & $17.5$ & $10.8$ & $12.2$ & $11.7$ & $0$ & $0$ & $3.7$ & $0.75$ & $0.5$ \\
{\MgI}$_{\rm 4}$ &  $12$ & $0.16$ & $-7.5$ & $200$ & $0.0005$ & $600$ & $17.5$ & $17.5$ & $10.8$ & $12.2$ & $11.7$ & $0$ & $0$ & $3.7$ & $0.75$ & $0.5$ \\
{\MgI}$_{\rm 1}$ & $24$ & $0.16$ & $-7.5$ & $200$ & $0.0005$ & $600$ & $17.5$ & $17.5$ & $10.8$ & $12.2$ & $11.7$ & $0$ & $0$ & $3.7$ & $0.75$ & $0.5$ \\

\hline
\multicolumn{15}{c}{\sc {\MgII}~Phase} \\
\hline
{\MgII}$_{\rm 1}$ & $-22$  & $0.16$ & $-3.7$ & $0.03$ & $60$ & $12000$ & $18.8$ & $17.3$ & $10.9$ & $13.1$ & $12.7$ & $11.8$ & $11.7$ & $16$ & $7.3$ & $7.0$ \\
{\MgII}$_{\rm 2}$ & $-10$   & $0.16$ & $-4.0$ & $0.06$ & $3$ & $11000$ & $17.8$ & $16.6$ & $10.5$ & $12.3$ & $12.0$ & $10.4$ & $10.1$ & $17$ & $10.4$ & $10.2$ \\
{\MgII}$_{\rm 3}$ & $1$  & $0.16$ & $-3.5$ & $0.02$ & $60$ & $13000$ & $18.6$ & $16.8$ & $10.7$ & $12.8$ & $12.1$ & $11.9$ & $12.0$ & $16$ & $6.5$ & $6.1$ \\
{\MgII}$_{\rm 4}$ & $12$ & $0.16$ & $-3.0$ & $0.006$ & $1000$ & $14000$ & $19.3$ & $17.0$ & $10.6$ & $13.0$ & $11.6$ & $13.1$ & $13.6$ & $21$ & $14.7$ & $14.5$ \\
{\MgII}$_{\rm 5}$ & $24$ & $0.16$ & $-2.5$ & $0.002$ & $500$ & $18000$ & $18.5$ & $15.7$ & $9.3$ & $11.5$ & $8.9$ & $12.6$ & $13.4$ & $18$ & $6.3$ & $5.7$ \\
{\MgII}$_{\rm 6}$ & $42$ & $0.16$ & $-2.5$ & $0.002$ & $500$ & $18000$ & $18.5$ & $15.7$ & $9.3$ & $11.5$ & $8.9$ & $12.6$ & $13.4$ & $17$ & $4.4$ & $3.6$ \\
{\MgII}$_{\rm 7}$ & $55$ & $0.16$ & $-2.7$ & $0.003$ & $50$ & $16000$ & $17.7$ & $15.1$ & $8.9$ & $11.0$ & $9.0$ & $11.7$ & $12.4$ & $17$ & $4.4$ & $3.6$ \\

\hline
\multicolumn{15}{c}{\sc Broad~{\HI}~Phase} \\
\hline
{\HI}$_{\rm 1}$ & $0$  & $0.003$ & $-1.5$ & $0.0002$ & $60000$ & $38000$ & $19.6$ & $15.5$ & $0$ & $7.3$ & $0$ & $9.5$ & $12.8$ & $58$ & $52.6$ & $52.5$ \\

\hline
\multicolumn{15}{c}{\sc Collisional~Ionization~Phase} \\
\hline
{\SiIV}$_{\rm 1}$ & $17$  & $0.16$ & \nodata & \nodata & \nodata & $63000$ & $19.2$ & $15.1$ & $0$ & $10.3$ & $12.0$ & $13.5$ & $13.1$ & $36$ & $17.2$ & $16.4$ \\

\hline
\enddata
\vglue -0.05in

\tablecomments{
\baselineskip=0.7\baselineskip
Column densities are listed in logarithmic units.}

\end{deluxetable}

\end{document}